\documentclass[11pt]{article}
\usepackage[DIV12]{typearea}
\usepackage{color}
\usepackage{amsmath,amssymb}
\usepackage{amsfonts}  
\usepackage{mathrsfs}
\usepackage{graphicx} % including PostScript
\usepackage{subfigure} 
\usepackage{bbm}
\usepackage{longtable}
\usepackage{subfigure}
\usepackage[small]{caption2} % font of captions is different from text
\usepackage{fleqn} % equations are not centered
\usepackage{cite}
\usepackage{pifont}
\usepackage{enumerate} % enumeration
\usepackage{pdflscape}
\usepackage{float}
\usepackage{multirow}
\usepackage{array}
\newcolumntype{C}{>$c<$}

\advance \headheight by 3.0truept       % for 12pt mandatory...
\DeclareMathOperator{\tr}{tr}

\newcommand{\E}[1]{\ensuremath{\mathrm{E}_{#1}}} % e.g. \E{8}

\newcommand{\SU}[1]{\ensuremath{\mathrm{SU}(#1)}}
\newcommand{\U}[1]{\ensuremath{\mathrm{U}(#1)}}
\newcommand{\Z}[1]{\ensuremath{\mathbbm{Z}_{#1}}} % Z_N ->\Z{N}

\newcommand{\bs}[1]{\ensuremath{\boldsymbol{#1}}}
\newcommand{\bsb}[1]{\ensuremath{\boldsymbol{\overline{#1}}}}
\newcommand{\maG}{\ensuremath{\mathcal{G}} }

\newcommand{\maN}{\ensuremath{\mathcal{N}} }

\newcommand{\x}{\ensuremath{\times}}

\allowdisplaybreaks

\begin{document}

%%%%%%%%%%%%%%%%%%%%%%%%%%%%%%%%%%%%%%%%%%%%%%%%%%%%%%%%%%%%%%%%%%%%%%%%%%
%  Title
%%%%%%%%%%%%%%%%%%%%%%%%%%%%%%%%%%%%%%%%%%%%%%%%%%%%%%%%%%%%%%%%%%%%%%%%%%
\date{}
\title{
  \vskip 2cm
  {\bf\huge $\boldsymbol{\U1'}$ coupling constant at low energies from heterotic orbifolds}\\[0.8cm]
}
\author{
 {\bf\normalsize
   Yessenia~Olgu\'in-Trejo,$^1$\footnote{\texttt{yes.olt@ciencias.unam.mx}}~
   Omar~P\'erez-Figueroa,$^1$\footnote{\texttt{omar\_perfig@ciencias.unam.mx}}~ 
   Ricardo~P\'erez-Mart\'inez$^{1,2}$\footnote{\texttt{ricardoperezm@estudiantes.fisica.unam.mx}}}\\ 
 {\bf\normalsize and Sa\'ul~Ramos-S\'anchez$^1$\footnote{\texttt{ramos@fisica.unam.mx}}
 }\\[1cm]
 {\it\normalsize $^1$Instituto de F\'isica, Universidad Nacional Aut\'onoma de M\'exico,}\\
 {\it\normalsize POB 20-364, Cd.Mx. 01000, M\'exico}\\[0.5cm]
{\it\normalsize $^2$Facultad de Ciencias F\'isico-Matem\'aticas, Universidad Aut\'onoma de Coahuila,}\\
 {\it\normalsize Edificio A, Unidad Camporredondo, 25000, Saltillo, Coahuila, M\'exico}
}

\maketitle 

\thispagestyle{empty}

\vskip 1cm
\begin{abstract}
Additional Abelian gauge interactions are generic to string compactifications. In
heterotic string models, gauge coupling unification of such forces and 
other gauge interactions is natural due to their common origin.
In this letter we study systematically the 1-loop running of the coupling constants in effective vacua
emerging from \Z8 heterotic orbifold compactifications that provide the matter spectrum of the
MSSM plus some vectorlike exotics, restricting to vacua that yield a non-anomalous $\U1'$ symmetry,
gauge coupling unification and the observed values of known gauge couplings.
We determine the low-energy value of the $\U1'$ coupling constant for different scales 
of supersymmetry breakdown.
We find that the $\U1'$ coupling constant is quite restricted in string models to lie in
the range 0.46--0.7 for low-scale supersymmetry or 0.44--0.6 in other cases.
We argue that the phenomenology of these string vacua should be further explored
to solve some extant issues, such as the stability of the Higgs vacuum.
\end{abstract}
\clearpage

\newpage

%%%%%%%%%%%%%%%%%%%%%%%%%%%%%%%%%%%%%%%%%%%%%%%%%%%%%%%%%%%%%%%%%%%%%%%%%%%%%%%%%%%%%%%%%%%%%%%%%%%%%%%%%
\section{Introduction}
\label{intro}

Some open questions in the standard model (SM) and cosmology have led to conjecture the existence of additional $\U1'$ gauge symmetries, 
under which different SM particles may be charged. These symmetries lead to a rich
phenomenology (for details, see ref.~\cite{Langacker:2008yv} and references therein). 
To mention a few of their qualities, they may shed some light 
on neutrino physics and dark matter simultaneously~\cite{Lebedev:2014bba,Rodrigues:2018jpv,Ellis:2018xal},
or the $g_\mu-2$ anomaly~\cite{Altmannshofer:2016brv},
or the metastability of the Higgs vacuum~\cite{DiChiara:2014wha}. 
They could also yield interesting signals at colliders~\cite{Martinez:2013qya,Celis:2015ara,Kim:2015hda} 
and alleviate some issues of models with supersymmetry (SUSY)~\cite{Erler:2000wu,Demir:2005ti}.
Although they must be broken at low energies and $m_{Z'}$ is very constrained~\cite{Aaboud:2017buh,Aaboud:2017sjh,Aaboud:2018mjh,Aaboud:2018tqo},
the bounds can be avoided (e.g. via a leptophobic $\U1'$) and $Z'$ signals could be soon confirmed at colliders.

The origin of $\U1'$ symmetries is frequently related to grand unified theories (GUT) beyond \SU5. For example, in
\E6 GUTs, $\U1'$ symmetries have been classified and studied phenomenologically~\cite{Erler:2011ud,Erler:2011iw,Rojas:2015tqa,Benavides:2018fzm}.
It is also known that $\U1'$s are natural to models resulting from different string 
compactifications~\cite{Komachenko:1989qn,Cvetic:2011iq,Cleaver:1997rk,Coriano:2007ba,Halverson:2018xge}. In particular,
in heterotic orbifolds in the fermionic formulation, plausible scenarios with a light
$Z'$ and rich phenomenology have been identified~\cite{Coriano:2007ba,Athanasopoulos:2014bba,Faraggi:2014ica}.
Also, orbifolds in the bosonic formulation have shown that in models resembling the minimal extension
of the SM (the MSSM), matter parity~\cite{Buchmuller:2006ik,Lebedev:2007hv} or even a $\Z4^R$ symmetry~\cite{Lee:2011dya} for proton
stability can arise from $\U1'$s (and other symmetries) of the model.

Motivated by these findings, in this letter we aim at characterizing the couplings 
of non-anomalous $\U1'$s natural to some string compactifications.
We focus on \Z8 toroidal orbifold compactifications of the \E8\x\E8 heterotic string with MSSM-like features.
This kind of models has been investigated before~\cite{Nibbelink:2013lua}, but with a different purpose and 
only in a small subset (about 20\%) of all promising models due to some far too restrictive priors 
which do not improve the phenomenology of the models. We avoid such restrictions to obtain a richer 
variety of models and a more general analysis.

We study special {\it effective vacua} with gauge group $\SU3_c\x\SU2_L\x\U1_Y\x\U1'$, where the kinetic mixing of Abelian
symmetries is unimportant~\cite{Goodsell:2011wn} and the few exotics in their spectra are vectorlike w.r.t. the SM.
Assuming that a Higgs-like mechanism breaks the $\U1'$ at a scale $\Lambda_{Z'}=2$ TeV, where also some
exotics acquire masses, and a SUSY breaking scale $\Lambda_{SUSY}\geq\Lambda_{Z'}$, we determine
systematically the running of all coupling constants of our effective vacua by using the
renormalization group equations (RGE) at 1-loop, which suffice at this level because of
further small corrections that we neglect, such as threshold effects. Restricting
to vacua with gauge coupling unification, we observe that besides the usual
family non-universality of stringy $\U1'$s, heterotic orbifolds limit the values 
of the $\U1'$ coupling at the TeV scale as well as the unification scale and the unified coupling.

In what follows, we discuss the main features of \Z8 heterotic orbifold compactifications
and their effective vacua with $\U1'$ and analyze how to arrive at limits for the couplings
of such symmetries. Our findings are reported in section~\ref{sec:results}, followed by 
a sample model with the potential to solve the metastability of the Higgs vacuum thanks to a $\U1'$.

%%%%%%%%%%%%%%%%%%%%%%%%%%%%%%%%%%%%%%%%%%%%%%%%%%%%%%%%%%%%%%%%%%%%%%%%%%%%%%%%%%%%%%%%%%%%%%%%%%%%%%%%%
\section{MSSM-like \bs{\Z8} toroidal orbifold models}
\label{sec:orbimodels}
%Orbifolds and \Z8 effective vacua

Our starting point is the $\maN=1$ \E8\x\E8 heterotic string theory in the bosonic formulation, and
we compactify the six extra dimensions on a toroidal \Z8 orbifold that preserves $\maN=1$ in four dimensions.
This choice is taken because \Z8 has shown to be the symmetry that yields the largest 
fraction of \Z{N} MSSM-like heterotic orbifold models available so far~\cite{Olguin-Trejo:2018wpw}, 
so that we can be sure to be focused on a representative patch of promising string compactifications.

In general, a toroidal \Z{N} orbifold is defined as the quotient $\mathbb{T}^6/P$ of a six-torus over a point group,
which is generated by a single twist $\vartheta$ of order $N$, i.e. so that $\vartheta^N=\mathbbm1$.
$\vartheta$ must be chosen to act as an isometry on $\mathbb{T}^6$. $\vartheta$ can always be
diagonalized on three two-dimensional actions, so that $\vartheta=\text{diag}(e^{2\pi i v_1}, e^{2\pi i v_2}, e^{2\pi i v_3})$,
where $v=(v_1,v_2,v_3)$ is called the {\it twist vector}. It is possible to combine together the point group
with the lattice $\Gamma$ of the torus to build the space group $S=P\ltimes\Gamma$, such that the orbifold
can be analogously defined as $\mathbbm{R}^6/S$.

A complete classification of the $\mathbb{T}^6$ geometries and point groups for Abelian toroidal orbifolds~\cite{Fischer:2012qj} 
reveals that there are only two \Z8 point groups, denoted \Z8--I and \Z8--II and defined by the twist vectors $v_{\Z8\text{--I}}=\frac18(1,2,-3)$
and $v_{\Z8\text{--II}}=\frac18(1,3,-4)$, respectively. There are five inequivalent $\mathbb{T}^6$ geometries (see~\cite{Nibbelink:2013lua} for details)
acceptable for these twists. Following the notation of~\cite{Fischer:2012qj}, we shall label them as \Z8--I ($i$,1), $i=1,2,3$, and \Z8--II ($j$,1), $j=1,2$.

A consistent heterotic orbifold requires to embed the action of the six-dimensional orbifold into
the \E8\x\E8 gauge degrees of freedom. The \Z{N} twist $\vartheta$ can be embedded as a 16-dimensional
{\it shift vector} $V$ of order $N$ (i.e. such that $NV$ is in the \E8\x\E8 root lattice), 
whereas the six independent directions of the torus can be embedded as 
16-dimensional discrete {\it Wilson lines} (WL) $W_a$, $a=1,\ldots,6$. Given a \Z{N} twist and a 
toroidal geometry, there are several admissible choices of shifts and WL, as 
long as they fulfill the {\it modular invariance conditions}~\cite{Ploger:2007iq},
\begin{eqnarray}
\label{eq:ModInv}
N\,(V^2-v^2) = 0\mod2\,, &\quad& N_a\,(V\cdot W_a) = 0\mod 2\,,\quad a=1,\ldots,6\,,\\
N_a\  W_a^2 = 0\mod 2\,, &\quad& \text{gcd}(N_a,N_b)\,(W_a\cdot W_b) = 0\mod 2\,,\quad a\neq b\,,\nonumber
\end{eqnarray}
which ensure that the four-dimensional emergent field theory be non-anomalous and compatible
with string theory. The space group of the orbifold constrains the order\footnote{The smallest integer $N_a$, 
such that $N_a W_a$ (with no summation over $a$) is contained in the root lattice of \E8\x\E8,
is defined as the {\it order of the WL} $W_a$.} $N_a$ of a WL
$W_a$ and its relations with other WL. Interestingly, these restrictions can be 
understood in terms of the Abelianization of the space group
as requirements for the WL to be compatible with the embedding of the so-called
space group flavor symmetry into the gauge degrees of freedom~\cite{Ramos-Sanchez:2018edc}.

In \Z8 orbifolds, the order and relations of the admissible WL are
given in table~\ref{tab:Z8WLConstraints}. Two \Z8--I geometries admit two independent WL
of order two, whereas the third case admits only one independent order-4 WL. Further, \Z8--II (1,1)
allows for three order-2 WL, and \Z8--II (2,1), one WL of order two and one of order four.

\begin{table}[t!]
\begin{center}
\begin{tabular}{|c|c|}
\hline
 Orbifold      & Conditions on the Wilson lines \\
\hline
\hline
 \Z8--I (1,1)  & $2W_1\approx 2W_5\approx 0;\;W_1\approx W_2\approx W_3\approx W_4;\, W_5\approx W_6$ \\
 \Z8--I (2,1)  & $2W_1\approx 2W_5\approx 0;\;W_1\approx W_2\approx W_3\approx W_4;\, W_5\approx W_6$ \\
 \Z8--I (3,1)  & $4W_1\approx 0;\;W_1\approx W_2\approx W_3\approx W_4\approx W_5\approx W_6$ \\
\hline
 \Z8--II (1,1) & $2W_1\approx2W_5\approx 2W_6\approx 0;\;W_1\approx W_2\approx W_3\approx W_4$ \\
 \Z8--II (2,1) & $4W_1\approx 2W_6\approx 0;\;W_1\approx W_2\approx W_3\approx W_4\approx W_5$ \\
\hline
\end{tabular}
\end{center}
\caption{Orders and relations of the WL of \Z8 orbifolds, depending on the geometry of the compactification.
$A\approx B$ indicates that $A=B$ up to translations in the root lattice of \E8\x\E8.}  
\label{tab:Z8WLConstraints}
\end{table}

Applying standard techniques (see e.g.~\cite{Dixon:1985jw,Dixon:1986jc,RamosSanchez:2008tn,Vaudrevange:2008sm}), 
one can use the modular invariant solutions to eqs.~\eqref{eq:ModInv} (with $N=8$) complying with the WL-constraints
of table~\ref{tab:Z8WLConstraints} to compute the spectrum of massless string states. These techniques
have been implemented in the \texttt{orbifolder}~\cite{Nilles:2011aj} to automatize the 
search of admissible models, the computation of the massless spectrum, and the identification of
phenomenologically viable models. By using this tool, we have previously~\cite{Olguin-Trejo:2018wpw} 
found 3,431 \Z8 orbifold compactifications of the \E8\x\E8 heterotic string with the following properties:

$\bullet$ the gauge group at the compactification scale is
          \begin{equation}
            \label{eq:G4D}
            \maG_{4D} = \maG_{SM}\x [\U1']^n \x \maG_\text{hidden}\,,
          \end{equation}
          where $\maG_{SM} = \SU3_c\x\SU2_L\x\U1_Y$ with non-anomalous hypercharge (satisfying $\sin^2\theta_w=3/8$), 
          $\maG_\text{hidden}$ is a non-Abelian gauge factor (typically a product of \SU{M} subgroups), at most 
          one $\U1'$ is (pseudo-)anomalous and $n\leq10$, depending on the model; and
        
$\bullet$ the massless spectrum includes the MSSM superfields plus
          only vectorlike exotic matter w.r.t. $\maG_{SM}$.

\noindent
$\maG_\text{hidden}$ is commonly considered a {\it hidden gauge group} because the MSSM fields
are mostly uncharged under that group.
The number of models found in each orbifold geometry is presented in table~\ref{tab:Z8MSSMs}.
The defining shifts and WL of the models can be found in~\cite{WebTablesFlavor:2018}.
     
Aiming at the study of the $\U1'$ symmetries of these models, we must point out some of their
properties. Most of the models present an anomalous $\U1'$~\cite{Witten:1984dg},
whose anomaly can be canceled through the Green-Schwarz mechanism~\cite{Green:1984sg}.
Besides, in this type of models, the gauge fields of the $\U1'_\alpha$ symmetries can be decomposed as
\begin{equation}
  \label{eq:gaugefields}
  T_\alpha = \sum_{I=1}^{16} t^I_\alpha H_I\,,\qquad \alpha=1,\ldots,n,
\end{equation}
in terms of the Cartan generators $H_I$ of the original \E8\x\E8, such that the corresponding
$\U1'_\alpha$ charges for fields of the spectrum with gauge momentum $p\in\E8\x\E8$
are given by $q_\alpha = t_\alpha \cdot p$. This is why $t_\alpha$ is frequently called the
generator of $\U1'_\alpha$. It is known that, if we adopt\footnote{Despite
this $\U1'$ normalization, we allow the GUT-compatible hypercharge normalization 
$|t_1|^2=5/6$.} the $\U1'$ normalization $k|t_\alpha|^2=1$ 
and consider all algebras associated with the gauge group to have Ka\v{c}-Moody level $k=2$, 
the tree-level gauge kinetic function is universally given by $f_\alpha = S$, where $S$ corresponds 
to the bosonic component of the (axio-)dilaton. Consequently, the tree-level gauge coupling 
satisfies $g_s^{-2} = \langle\text{Re} S\rangle$ at the (heterotic) string scale, $M_{str}\approx 10^{17}$ GeV.
Further, there is some kinetic mixing between different $\U1'$ gauge symmetries which one might believe
relevant for phenomenology; however, it has been found to be typically of order $10^{-4}$--$10^{-2}$ in semi-realistic heterotic 
orbifolds~\cite{Goodsell:2011wn}, unimportant for our purposes.

\begin{table}[t!]
\begin{center}
\begin{tabular}{|c|c|c||c|c|c|}
\hline 
Orbifold                  & \# MSSM-like & effective & Orbifold                    & \# MSSM-like & effective\\[-2mm]
                          & models       & vacua     &                             & models       & vacua    \\
\hline 
$\Z{8}\text{--I}\,(1,1)$  &  268         & 1,362     & $\Z{8}\text{--II}\,(1,1)$   & 2,023        & 10,023   \\ 
$\Z{8}\text{--I}\,(2,1)$  &  246         & 1,097     & $\Z{8}\text{--II}\,(2,1)$   &   505        &  2,813   \\ 
$\Z{8}\text{--I}\,(3,1)$  &  389         & 1,989     &                             &              &          \\ 
\hline
\end{tabular}
\caption{Number of inequivalent heterotic orbifold models with MSSM-like properties found 
         in ref.~\cite{Olguin-Trejo:2018wpw} for each \Z8 orbifold geometry. We further provide 
         in the third and sixth columns the number of vacuum configurations with MSSM-like properties and gauge group
         $\maG_\text{eff}=\maG_{SM}\x\U1'$ in each case.} 
\label{tab:Z8MSSMs}
\end{center}
\end{table}

Let us make a couple of additional remarks about the models we explore. First, 
as in the MiniLandscape~\cite{Lebedev:2006kn,Lebedev:2007hv,Lebedev:2008un}
of \Z6--II heterotic orbifolds, in all the \Z8 models discussed here, there exists a large number of SM-singlets, 
which naturally develop $\mathcal{O}(0.1)$ VEVs in order to cancel the Fayet-Iliopoulos term, $\xi= g_s^2 \tr T_\text{anom}/192\pi^2$ (in Planck units), 
appearing in models with a (pseudo-)anomalous $\U1'$. As a consequence, the allowed couplings of such singlets among themselves and with 
vectorlike exotics yield large masses for the additional matter, decoupling it from
the low-energy effective field theory. Simultaneously, since the singlets are charged under
the $[\U1']^n$ gauge sector, those symmetries can be broken in the vacuum. 

Secondly, it is always possible to find SM-singlet VEV configurations, such that, SUSY is
retained while only the {\it effective gauge group},
\begin{equation}
 \label{eq:Geff}
   \maG_\text{eff} = \maG_{SM}\x \U1' \subset \maG_{4D}\,,
\end{equation}
remains after the spontaneous breakdown triggered by the singlet VEVs,
where we ignore the hidden group $\maG_\text{hidden}$.
The surviving (non-anomalous) $\U1'$ can be any of the original $\U1'_\alpha$, $\alpha=1,\ldots,n$, 
symmetries or a linear combination of them, depending on the details of the model.
In this work, for practical purposes, we shall study only the former case, i.e. the
effective vacua with the effective gauge group $\maG_\text{eff}$, where the $\U1'$ corresponds
to each of the non-anomalous $\U1'_\alpha$ of the \Z8 orbifold models. We find
that there is a total of 17,284 such effective vacua, distributed in all admissible
\Z8 orbifold geometries, as shown in table~\ref{tab:Z8MSSMs}.

Third, in the most general case, (at least) some of the MSSM superfields and the 
exotics exhibit some $\U1'$ charges. This is true in most of the vacua where one of the 
$\U1'$ remains unbroken. Thus, only the exotics that are vectorlike w.r.t. $\maG_\text{eff}$, 
and not just $\maG_{SM}$, decouple at low energies, allowing interactions among SM fields
charged under $\U1'$ and the extra matter which can yield interesting new phenomenology.
These interactions affect in particular the RGE running of the gauge couplings, as we discuss in the 
following section.

%%%%%%%%%%%%%%%%%%%%%%%%%%%%%%%%%%%%%%%%%%%%%%%%%%%%%%%%%%%%%%%%%%%%%%%%%%%%%%%%%%%%%%%%%%%%%%%%%%%%%%%%%
\section{Searching effective vacua with $\bs{\U1'}$ and unification}
\label{sec:procedure}
%Method to arrive at our results

We shall study the value of the $\U1'$ coupling constants of the effective field theories emerging from \Z8 heterotic orbifold 
compactifications at currently reachable energies, restricting ourselves to those that are consistent with unification,
in the sense that the gauge couplings meet (perhaps accidentally) at a given scale,
and where SM gauge couplings are compatible with the observed values at low energies. 

Selecting the vacua with these features requires some additional knowledge of the details of the effective spectrum,
and some reasonable priors. The first hurdle is that, even though $g_s^{-2}=\langle \text{Re} S\rangle$ at $M_{str}$,
below this scale, the coupling constants get different contributions from all other moduli too,
whose stabilization represents a challenge by itself~\cite{Parameswaran:2010ec,Dundee:2010sb},
hindering to know the exact values of the gauge couplings in the effective field theory.
Thus, in order to figure out which models yield the measured values of the gauge couplings, 
one could start with an {\it ad hoc} value of all SM gauge couplings at high energies and 
retain only models where the RGE lead to the observed values at $M_Z$. This approach
seems rather arbitrary. 

Instead, we assume that the SM coupling strengths $\alpha_i$ of our effective vacua have 
the observed values at $M_Z$~\cite{Patrignani:2016xqp},
(with $i=1$ for $\U1_Y$, $i=2$ for $\SU2_L$ and $i=3$ for $\SU3_c$)
\begin{equation}
\label{eq:alphasMZ}
\alpha^{-1}_1(M_Z) = 59.01\pm 0.01\,,\quad
\alpha^{-1}_2(M_Z) = 29.59\pm 0.01\,,\quad
\alpha_3(M_Z) = 0.1182\pm 0.0012\,,
\end{equation}
and then
let the RGE define the value of the gauge couplings at all scales up to $M_{str}$, using the spectrum of the effective vacua
we find. These effective vacua exhibit $\maN=1$ SUSY and an additional $Z'$ boson, but no SUSY partner
(in some models, the lightest neutralino with masses lighter than few hundred GeV has been excluded~\cite{Sirunyan:2018ell,Aaboud:2018ngk})
nor extra vector boson has been detected at the LHC so far (with a lower limit for $m_{Z'}$ around 
2 TeV~\cite{Aaboud:2017buh,Aaboud:2017sjh,Aaboud:2018mjh,Aaboud:2018tqo}). Hence, we suppose that
SUSY is broken at a scale $\Lambda_{SUSY}>M_Z$ and the $\U1'$ breakdown scale is $\Lambda_{Z'}=2$ TeV, 
as a benchmark value.\footnote{We have verified that our results are qualitatively the same for other scales 
near this value.} We further assume that $\Lambda_{SUSY} \geq \Lambda_{Z'}$.

\begin{figure}[!t!]
\begin{center}
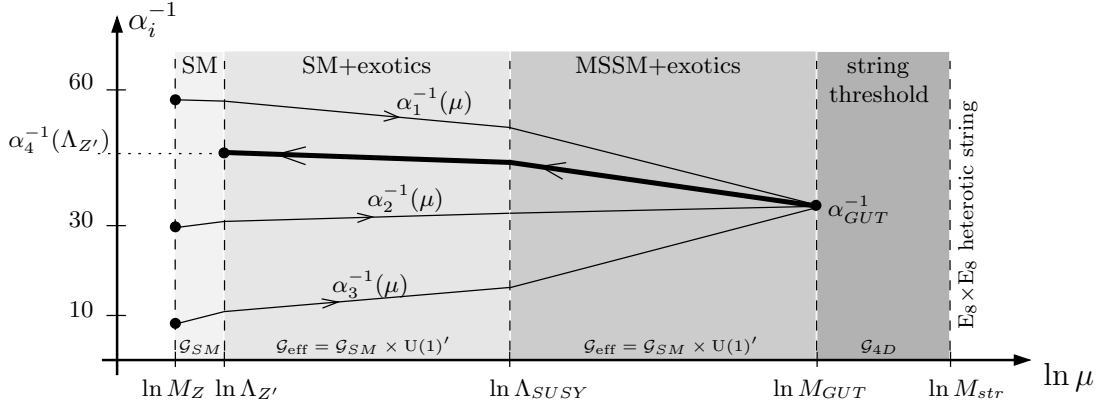
\caption{Our approach to determine the value at low energies of the $\U1'$ coupling
strength $\alpha_4$, the unification scale $M_{GUT}$ and the unification value of the gauge couplings $\alpha_{GUT}$.
We consider $\Lambda_{Z'}=2$ TeV, the three cases $\Lambda_{SUSY}=\Lambda_{Z'}, 10^{12}$ GeV$, 10^{17}$ GeV
and $M_{str}\approx10^{17}$ GeV. We illustrate the case $\Lambda_{Z'}<\Lambda_{SUSY}<M_{GUT}$.}
\label{fig:approach}
\end{center}
\end{figure}

Given these remarks, we consider different matter spectra depending on the energy scale $\mu$.
As sketched in figure~\ref{fig:approach}, above the SUSY scale $\Lambda_{SUSY}$ the spectrum of the effective
vacua includes the MSSM superfields and a few vectorlike exotics w.r.t. $\maG_{SM}$ with non-trivial 
$\U1'$ charges. Below $\Lambda_{SUSY}$, if it is larger than $\Lambda_{Z'}$, the gauge group is still
$\maG_\text{eff}$, eq.~\eqref{eq:Geff}, but we assume that all SM superpartners and the bosonic superpartners of the
exotics decouple. Only one SM singlet with $\U1'$ charge is taken as scalar below $\Lambda_{SUSY}$, so that its VEV
can trigger the breakdown of $\U1'$ and provide masses around $\Lambda_{Z'}$ for the remaining exotics.
Consequenly, below $\Lambda_{Z'}$ only the SM particles and gauge group are left.

The next step is to choose only models that are consistent with gauge coupling unification, so that we
recover at some level the unification provided at $M_{str}$ by string theory, and justify why we have 
restricted ourselves to hypercharges with GUT normalization. To do so, we let the RGE determine the 
value of the SM couplings and retain vacua with unification at some scale $M_{GUT}$, which varies from 
vacuum to vacuum due to their matter content.\footnote{Note that, even though $M_{GUT}$ defines the scale
at which all gauge couplings meet, no true unification is implied in the sense of usual GUTs.} 
At $M_{GUT}$, one naturally can assume that all coupling 
strengths of $\maG_\text{eff}$ have the same value $\alpha_{GUT}$. Thus, the $\U1'$ coupling strength 
$\alpha_4=\alpha_{GUT}$ at that scale and its RGE running down to $\Lambda_{Z'}$ in a vacuum 
is a low-energy consequence of such a vacuum. This approach is depicted in figure~\ref{fig:approach}.
We display an intermediate $\Lambda_{SUSY}$, but, in this work, we consider three well-motivated cases: $\Lambda_{SUSY} = \Lambda_{Z'}$,
$\Lambda_{SUSY} = 10^{12}$ GeV and $\Lambda_{SUSY} = 10^{17}$ GeV.
The first one arises from the common expectation that SUSY may show up at reachable energies,
the second one from constraints on the metastability of the Higgs potential~\cite{Ibanez:2013gf} 
and the last one from considering that SUSY may be broken at the string scale, as in non-SUSY string 
compactifications~\cite{Blaszczyk:2014qoa,Nibbelink:2015vha,Abel:2018zyt}.
It is known that gaugino condensates can render various $\Lambda_{SUSY}$ in these models~\cite{Lebedev:2006tr}.

As stated before, all vectorlike exotics w.r.t. $\maG_\text{eff}$ naturally acquire masses just below $M_{str}$
while $\maG_{4D}$ breaks down to $\maG_\text{eff}$, but it is easy to conceive that this process happens gradually 
at various scales between $M_{GUT}$ and $M_{str}$. Furthermore, we expect that at those scales string threshold corrections,
effects of the Green-Schwarz mechanism and moduli stabilization take place. We shall assume that all these effects do not alter 
the unification, even though they set deviations of $\alpha_{GUT}$ from $\alpha_s=g_s^2/4\pi$ that differ for
each effective vacuum.

In this work, we use the 1-loop RGE for the gauge factors of the effective gauge group, $G_i\in\maG_\text{eff}$.
The running of the coupling strengths $\alpha_i = g_i^2/4\pi$ is given by
\begin{equation}
\label{eq:running}
 \frac{\partial \alpha_i^{-1}}{\partial \ln\mu} = - \frac{b_i}{2\pi}\,,\qquad i=1,\ldots,4\,,
\end{equation}
where the $\beta$-function coefficients for non-Abelian groups  ($i=2,3$) are given at 1-loop by 
\begin{equation}
\label{eq:nonAbelianb}
   b_i= \left\{
   \begin{array}{lcl}
      -\frac{11}{3}C_2(G_i)+\frac23 \sum_f m_f C(\bs R_f) + \frac13 \sum_b m_b C(\bs R_b)\,,&\quad& \text{non-SUSY}\,,\\
      -3\, C_2(G_i) + \sum_S m_S\, C(\bs R_S) \,,&\quad& \text{SUSY}\,,\\
   \end{array}
   \right.
\end{equation}
Here $C_2(G_i)$ is the quadratic Casimir of the group $G_i$, $C(\bs R_{b,f,S})$ denotes respectively the quadratic index
of the $G_i$ representations $\bs R_{b,f,S}$ of the bosons, fermions and superfields included in the spectrum,
over which the sums run, and $m_{b,f,S}$ denotes their multiplicities. Conventionally, we take $C(\bs R_{b,f,S})=1/2$ if $\bs R_{b,f,S}$
corresponds to the fundamental representation of \SU{N}. For Abelian gauge symmetries ($i=1,4$), eq.~\eqref{eq:nonAbelianb}
reduces to
\begin{equation}
\label{eq:Abelianb}
   b_i= \left\{
   \begin{array}{lcl}
      \frac23 \sum_f m_f |q_i^{(f)}|^2 + \frac13 \sum_b m_b |q_i^{(b)}|^2\,,&\quad& \text{non-SUSY}\,,\\
      \sum_S m_S\, \tr |q_i^{(S)}|^2 \,,&\quad& \text{SUSY}\,.\\
   \end{array}
   \right.
\end{equation}
in terms of the matter $\U1'$ charges $q_i^{(b,f,S)}$, defined around eq.~\eqref{eq:gaugefields}. 

The solutions to~\eqref{eq:running} have the general form 
$\alpha_i^{-1}(\mu) = \alpha_i^{-1}(\mu_0) - b_i/2\pi \ln(\mu/\mu_0)$, in terms of a 
reference scale $\mu_0$. For the SM couplings, we take their observed 
values, eq.~\eqref{eq:alphasMZ}, with $\mu_0 = M_Z$. Since below $\Lambda_{Z'}=2$ TeV 
we assume that only the SM particles are present, we determine readily that
$\alpha^{-1}_1 = 56.99$, $\alpha^{-1}_2 = 31.15$ and $\alpha^{-1}_3 = 11.9$
at $\Lambda_{Z'}$.

%%%%%%%%%%%%%%%%%%%%%%%%%%%%%%%%%%%%%%%%%%%%%%%%%%%%%%%%%%%%%%%%%%%%%%%%%%%%%%%%%%%%%%%%%%%%%%%%%%%%%%%%%
\section{$\bs{\U1'}$ couplings in \bs{\Z8} heterotic orbifold vacua}
\label{sec:results}
%Our main results

As a first step, we compute systematically the $\beta$-function coefficients, according to eqs.~\eqref{eq:nonAbelianb}-\eqref{eq:Abelianb},
of all effective vacua counted in table~\ref{tab:Z8MSSMs}. We observe that a fraction (3.5\%) of all vacua yield the
properties of the MSSM above $\Lambda_{SUSY}$. That is, their matter spectra match the MSSM spectrum and, consequently,
$(b_1,b_2,b_3,b_4)=(33/5,1,-3,0)$. In these cases, $b_4=0$ arises because the MSSM fields have no $\U1'$ charges.
In the second column of table~\ref{tab:analyzedvacua} we show the number of these MSSM vacua.
There is also a smaller fraction (1.3\%) of vacua with $b_4=0$, but $b_i\neq b_i^{MSSM}$.
Since in all these cases the coupling of $Z'$ with observable matter is very suppressed, we shall not consider 
these models here.

The running of the couplings described by the RGE reveals that there is a significant number of vacua that are 
inadmissible for our study. First, computing the scale at which couplings meet leads in some cases to $M_{GUT}<M_Z$ or $M_{GUT}>M_{str}$,
which are either excluded (the former) or meaningless since our effective models apply only below $M_{str}$.
Second, vacua in which any of the coupling strengths $\alpha_i=g_i^2/4\pi$ reaches negative values in its running
are not acceptable. Finally, since our work is based on weakly coupled string theory, non-perturbative couplings,
$\alpha_i>1$, are equally undesirable.

The number of vacua with these weaknesses varies depending on the choice of $\Lambda_{SUSY}$ and the \Z8 orbifold geometry.
For instance, in the case of \Z8-I (1,1) with $\Lambda_{SUSY}=\Lambda_{Z'}=2$ TeV, we find that out of the 1,362 effective vacua,
205 lead to unification at a scale $M_{GUT}<\Lambda_{Z'}$ or $M_{GUT}>M_{str}$. Further, 321 vacua produce negative 
values of some $\alpha_i$, and in 89 we find non-perturbative values for some couplings ($\alpha_i >1$).
Disregarding these (and those with $b_4=0$), we arrive at the 681 vacua of table~\ref{tab:Z8MSSMs} for this case.

To obtain the \Z8 orbifold vacua of interest, with unification of all coupling strengths, we proceed in two steps. We analyze first
the qualities of those vacua with $\SU2_L-\U1_Y$ unification, i.e. those with $\alpha_1=\alpha_2\neq\alpha_3$, and then
select among them those with $\alpha_{GUT}\equiv\alpha_1 = \alpha_2\approx\alpha_3$ at a scale $M_{GUT}$, allowing for
a small deviation $|\alpha_{GUT}^{-1}-\alpha_3^{-1}(M_{GUT})|<0.26$, corresponding to the $3\sigma$ interval of the 
measured value of $\alpha_3^{-1}(M_Z)$. These observations are considered in the third through eighth columns of
table~\ref{tab:analyzedvacua}, where we display the number of vacua with partial and full unification for each
choice of $\Lambda_{SUSY}$. We realize that, from the huge number of possible effective vacua with $\maG_\text{eff}$,
only a quite small set of \Z8 orbifold vacua of order hundred in each case satisfies all of our constraints. 
The details of these vacua are available in~\cite{WebTables:2019z}.

\begin{table}[t!]
\begin{center}
{
\begin{tabular}{|c|c|cc|cc|cc|}
\hline 
\multirow{3}{*}{orbifold} & \multirow{3}{*}{$\begin{array}{C}MSSM\\vacua\end{array}$} 
                                         & \multicolumn{6}{c|}{effective vacua of interest}          \\ 
                          &
                                         & \multicolumn{2}{c|}{$\Lambda_{SUSY}=\Lambda_{Z'}$} & \multicolumn{2}{c|}{$\Lambda_{SUSY}=10^{12}$ GeV} & \multicolumn{2}{c|}{$\Lambda_{SUSY}=10^{17}$ GeV} \\
                          &
                                         & $\alpha_1=\alpha_2$ & unified & $\alpha_1=\alpha_2$ & unified & $\alpha_1=\alpha_2$ & unified \\
\hline
\Z8--I (1,1)           &   58            &     681             &   2     &  1,218              &   8     &  1,253              &   20\\
\hline  
\Z8--I (2,1)           &  116            &     421             &   0     &    940              &   2     &    958              &    3\\
\hline
\Z8--I (3,1)           &   76            &   1,101             &   8     &  1,792              &   7     &  1,844              &   20\\
\hline
\Z8--II (1,1)          &  245            &   6,476             &  60     &  8,970              & 181     &  9,245              &  114\\
\hline
\Z8--II (2,1)          &  111            &   1,929             &   7     &  2,567              &  71     &  2,631              &   51\\
\hline\hline
totals                 &  606            &  10,608             &  77     & 15,487              & 269     & 15,931              &  208\\
\hline
\end{tabular}
\caption{Number of effective MSSM-like vacua arising from \Z8 heterotic orbifolds with $\U1'$ symmetries 
         and gauge coupling unification. For each orbifold geometry, the second column shows the number
         of vacua with the exact dynamics of the MSSM, i.e. such that $b_i=b_i^{MSSM}\in\{33/5,1,-3,0\}$
         in their RGE. Excluding these and other inconsistent models, in the remaining (pairs of) columns we show the number of vacua 
         satisfying our constraints and with partial unification ($\alpha_1=\alpha_2$) or total
         gauge coupling unification ($\alpha_{GUT}=\alpha_i$), corresponding to three choices of $\Lambda_{SUSY}$.}
\label{tab:analyzedvacua}
}
\end{center}
\end{table}

\begin{figure}[t!]
\centering
\subfigure{\includegraphics[width=0.49\textwidth]
{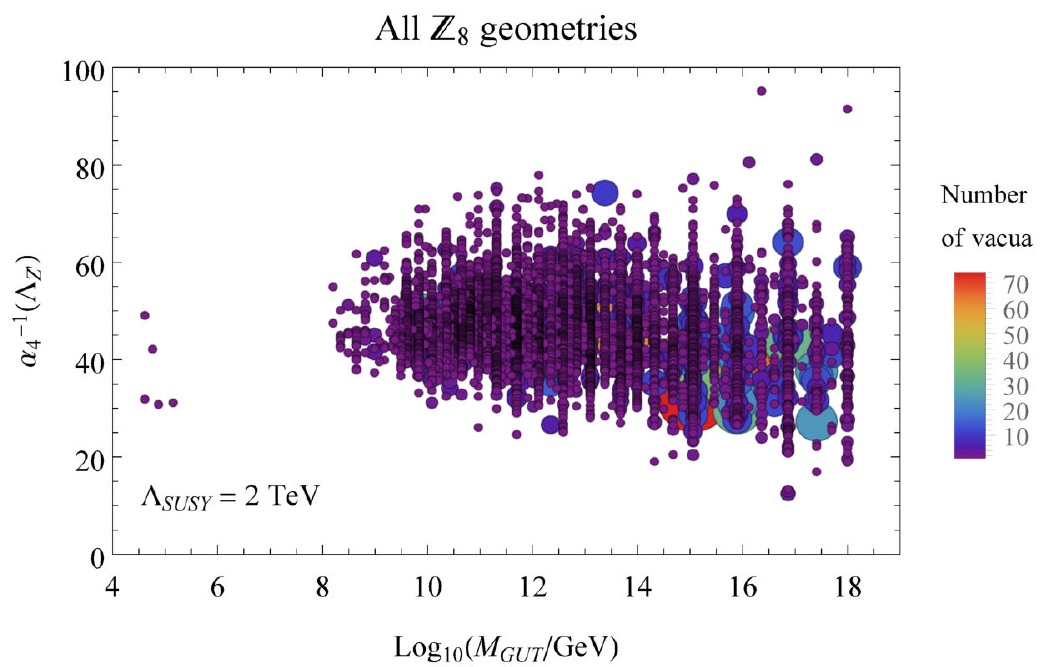}}
\subfigure{\includegraphics[width=0.49\textwidth]
{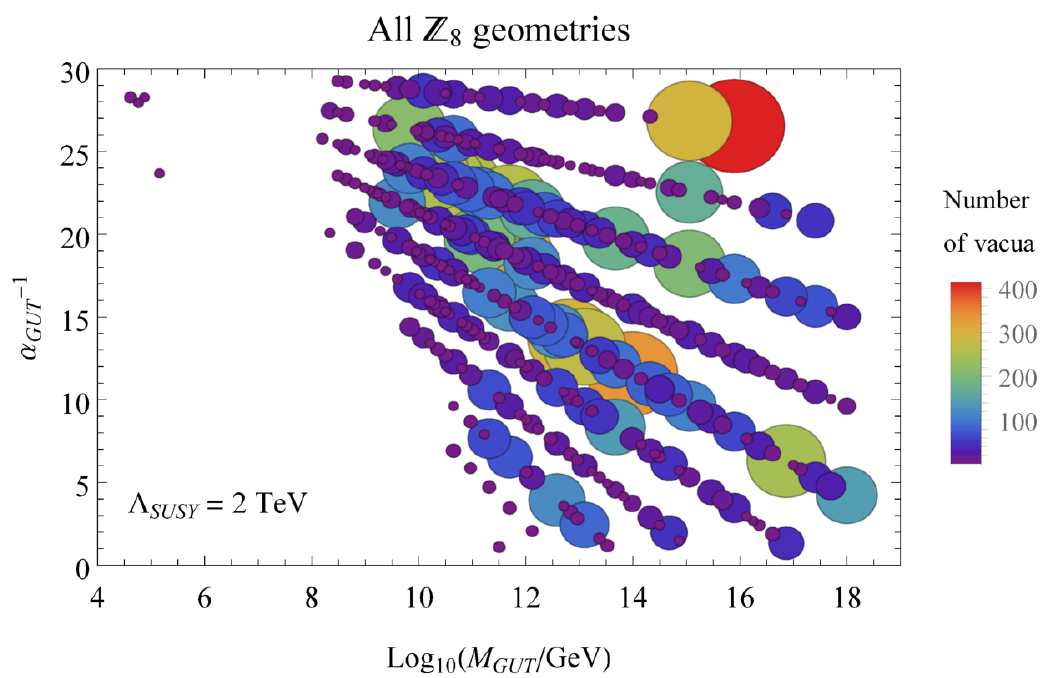}}

\subfigure{\includegraphics[width=0.49\textwidth]
{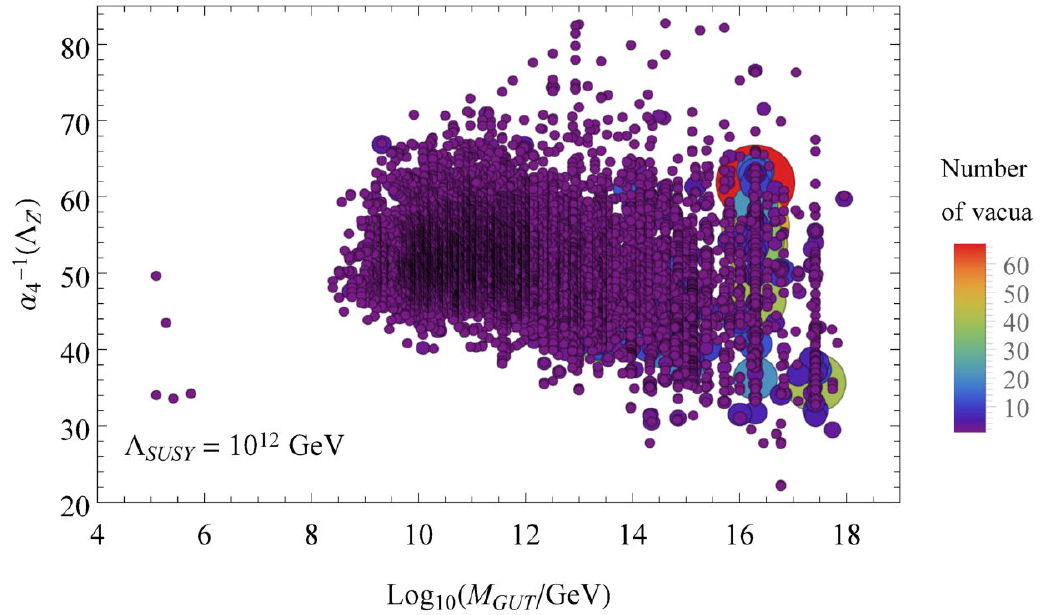}}
\subfigure{\includegraphics[width=0.49\textwidth]
{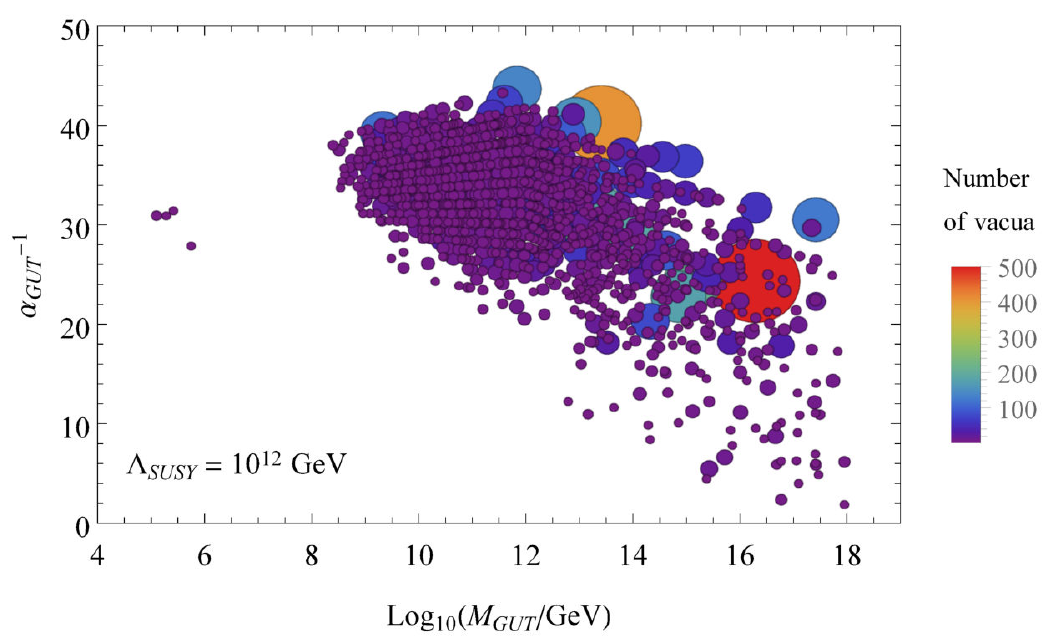}}

\subfigure{\includegraphics[width=0.49\textwidth]
{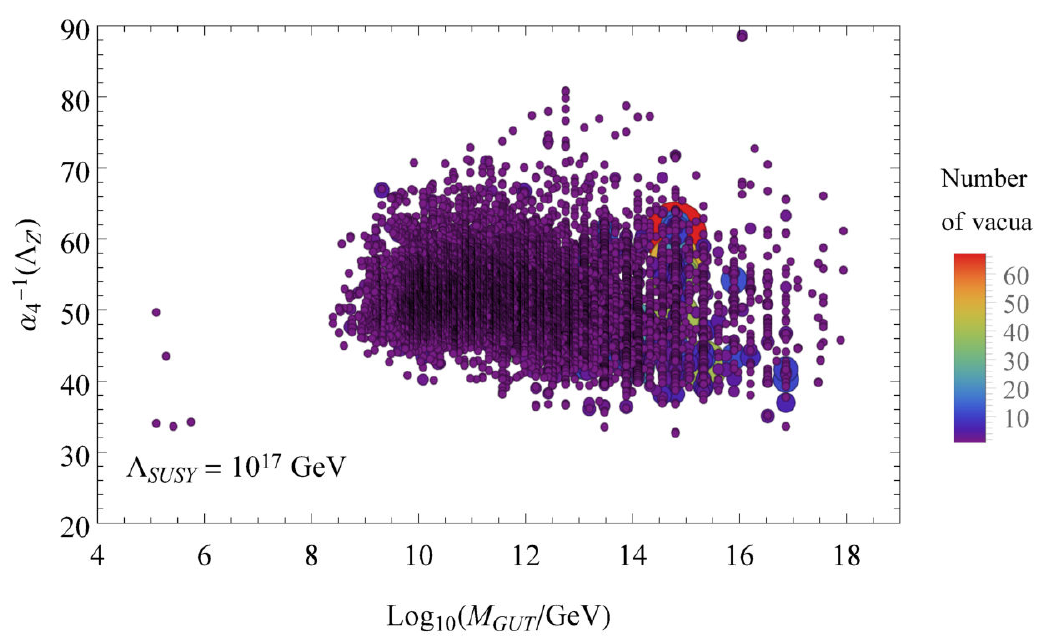}}
\subfigure{\includegraphics[width=0.49\textwidth]
{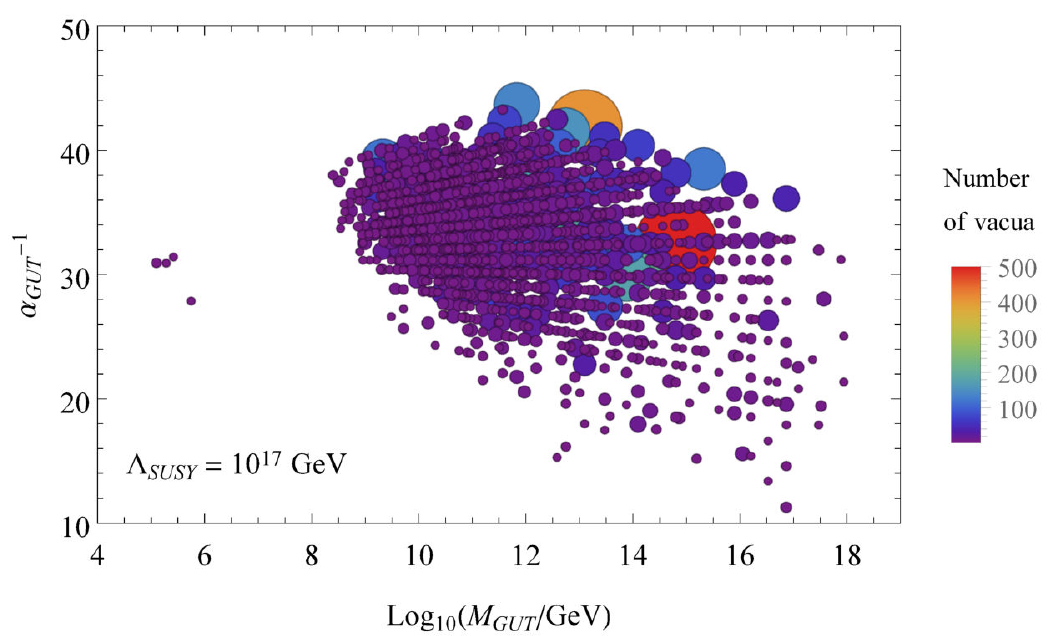}}

\caption{Vacua with different values of $M_{GUT}$, $\alpha_4^{-1}(\Lambda_{Z'})$ and $\alpha_{GUT}^{-1}$
         for three choices of $\Lambda_{SUSY}$ and partial unification, $\alpha_{GUT}\equiv\alpha_1=\alpha_2\neq\alpha_3$. 
         In the left panels, the bubbles in different colors and sizes indicate the number of vacua
         with the given values of $\alpha_4(\Lambda_{Z'})$ and $M_{GUT}$ at their center. Analogously,
         the right panels count vacua with different values of $\alpha_{GUT}$ and $M_{GUT}$.} 
\label{fig:Z8allgeom}
\end{figure}

In figure~\ref{fig:Z8allgeom} we present our results for all \Z8 orbifold vacua with only $\SU2_L-\U1_Y$
unification. The left panels correspond to frequency plots for three different choices of $\Lambda_{SUSY}$ 
of (the inverse of) the $\U1'$ coupling strength $\alpha_4^{-1}$
at the low-energy scale $\Lambda_{Z'}=2$ TeV against the scale at which $\alpha_1=\alpha_2$, denoted $M_{GUT}$.
The central values of the largest (red) bubble corresponds to the most frequent combination of $\alpha_4^{-1}(\Lambda_{Z'})$ and $M_{GUT}$.
The small (purple) bubbles correspond to at most six vacua with the combination of values at their
center. The right panels are also frequency plots of the values of $\alpha_{GUT}^{-1}$ and $M_{GUT}$
achieved by our vacua, where $\alpha_{GUT}\equiv\alpha_1(M_{GUT})=\alpha_2(M_{GUT})$. In these plots,
small purple bubbles correspond to up to 50 models with the central value of the circles.

Since this is only an intermediate result, we content ourselves with some semi-qualitative remarks.
The first observation is that, independently of whether SUSY is broken at low, intermediate or high energies,
\Z8 orbifold vacua with MSSM-like properties do not allow any arbitrary values of $\U1'$ couplings constants
or unification scale. We find that roughly only $20 < \alpha_4^{-1}(2\text{ TeV}) < 80$, corresponding to
$0.4 < g_4(2\text{ TeV}) < 0.8$, is allowed in our string constructions. We expect this to hold
for any heterotic orbifold model with semi-realistic properties. The most common value of $\alpha_4^{-1}(2\text{ TeV})$
depends on $\Lambda_{SUSY}$: for low-scale SUSY $\alpha_4^{-1}\sim 30$, whereas $\alpha_4^{-1}\sim 60$
for other cases. We observe also other rough limits: $M_{GUT}>10^{8}\text{ GeV}$, and $\alpha_{GUT}^{-1}<30$ for 
low-scale SUSY and $\alpha_{GUT}^{-1}<45$ for other SUSY scales.

On the other hand, taking averages over all models, we find $\overline{M_{GUT}}\approx10^{16}$ GeV and 
$\overline{\alpha_{GUT}^{-1}}\approx 13$ for low-scale SUSY, and 
$\overline{M_{GUT}}\approx10^{15}$ GeV and $\overline{\alpha_{GUT}^{-1}}\approx 30$ otherwise.
Unfortunately, most of these vacua are far from our ideal scenario, with full unification, which can also be measured by 
the average difference of $\Delta\equiv|\alpha_{GUT}-\alpha_3(M_{GUT})|$, which is as large as $\overline\Delta\approx\alpha_{GUT}$
for low-scale SUSY, and $\overline\Delta\approx\frac13\alpha_{GUT}$ for higher $\Lambda_{SUSY}$.

% Our data yields
%$\Lambda_{SUSY}=\Lambda_{Z'} \to  |\Delta \alpha|= 0.0692   \to   M_{GUT}= 2.52824\x10^{16} \text{ GeV}\to  \alpha_{GUT}=0.07966$
%$\Lambda_{SUSY}=10^{12} \text{  GeV} \to   |\Delta \alpha|= 0.0122   \to   M_{GUT}= 4.94889\x10^{15} \text{ GeV} \to   \alpha_{GUT}=0.03174$
%$\Lambda_{SUSY}=10^{17} \text{ GeV} \to   |\Delta\alpha|= 0.0108  \to   M_{GUT}= 1.11785\x10^{15} \text{ GeV} \to  \alpha_{GUT}= 0.02954$

Let us comment on the vertical (diagonal) alignment pattern of the points in the left (right) plots 
of figure~\ref{fig:Z8allgeom}. Consider e.g. the top plots, with $\Lambda_{SUSY}=\Lambda_{Z'}$, where one can find that the RGE lead to
\begin{equation*}
\ln\frac{M_{GUT}}{\Lambda_{Z'}} = 2\pi\left(\alpha_1^{-1}(\Lambda_{Z'})-\alpha_2^{-1}(\Lambda_{Z'})\right) \frac{1}{b_1-b_2}\,.
\end{equation*}
Since only $M_{GUT}$ and $b_1-b_2$ are model-dependent, all vacua with the same difference $b_1-b_2$
lead to the same $M_{GUT}$, yielding a point on the same vertical line in the left panels of the figure. 
Further, as not any arbitrary $b_i$ can appear in our vacua (but only rational numbers), this difference
does not build a continuous, producing separate lines. The origin of the diagonal lines in the right
panels is simpler. The RGE lead to 
$\alpha_{GUT}^{-1} = \alpha_2^{-1}(M_{GUT}) = \alpha_2^{-1}(\Lambda_{Z'})-\frac{b_2}{2\pi}\ln\frac{M_{GUT}}{\Lambda_{Z'}}$,
as a result of the running of $\alpha_2$; each diagonal line describes this running for a given $b_2$, populated by all 
vacua with the same $b_2$.

\begin{figure}[t!]
\centering

\subfigure{\includegraphics[width=0.49\textwidth]
{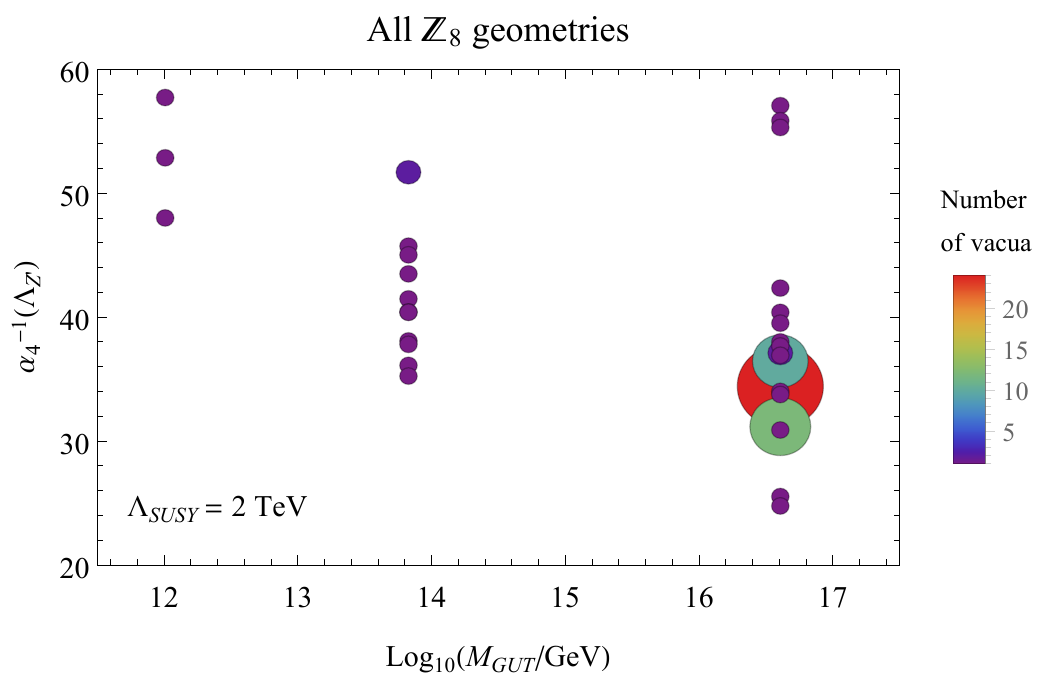}}
\subfigure{\includegraphics[width=0.49\textwidth]
{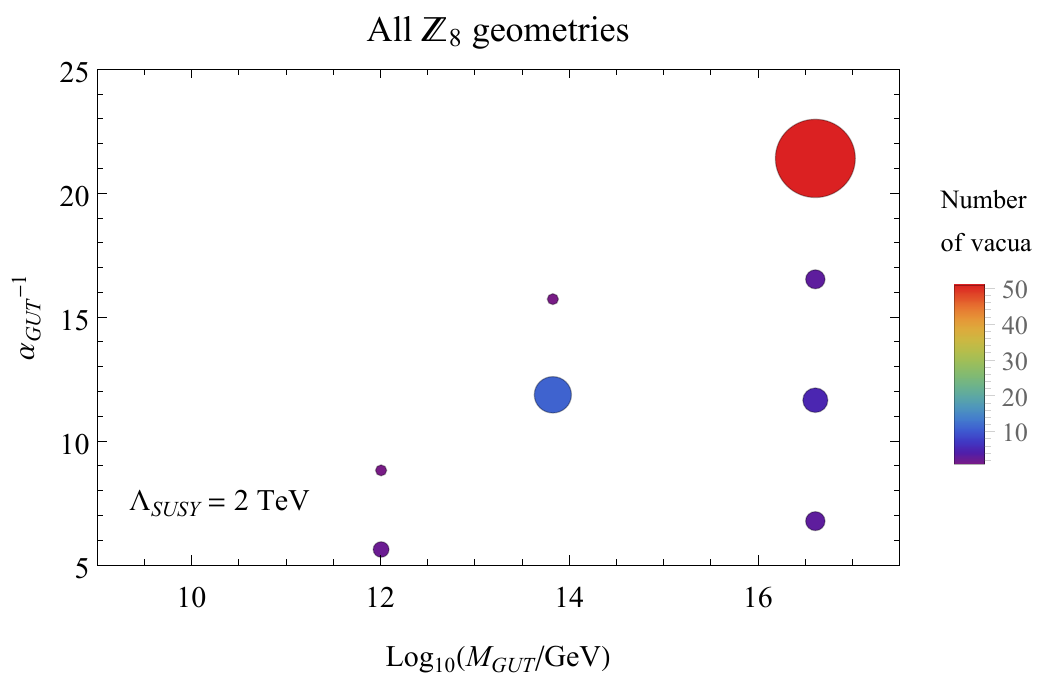}}

\subfigure{\includegraphics[width=0.49\textwidth]
{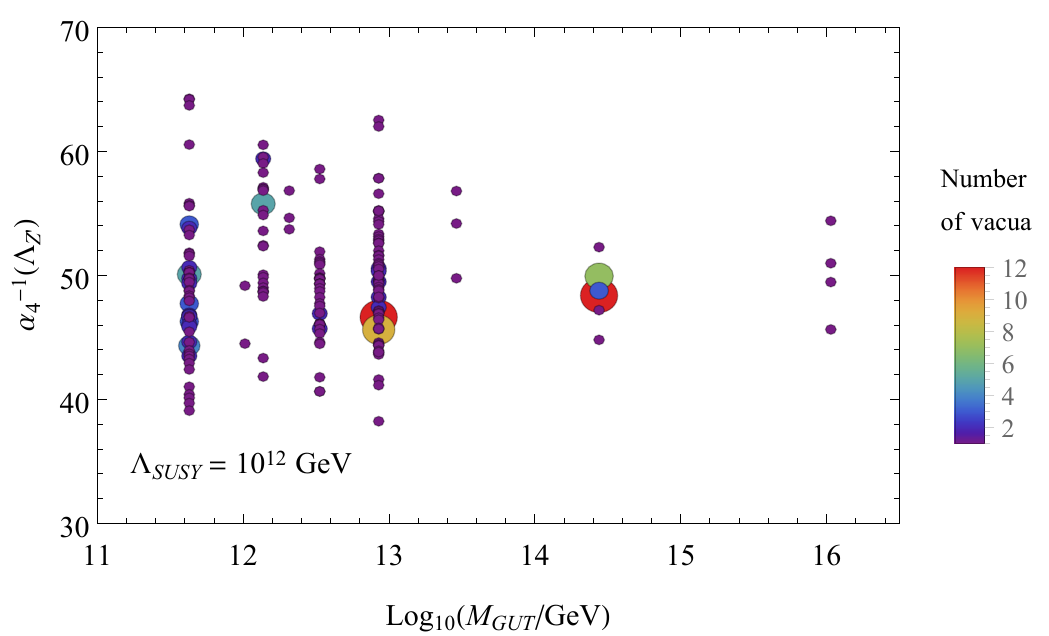}}
\subfigure{\includegraphics[width=0.49\textwidth]
{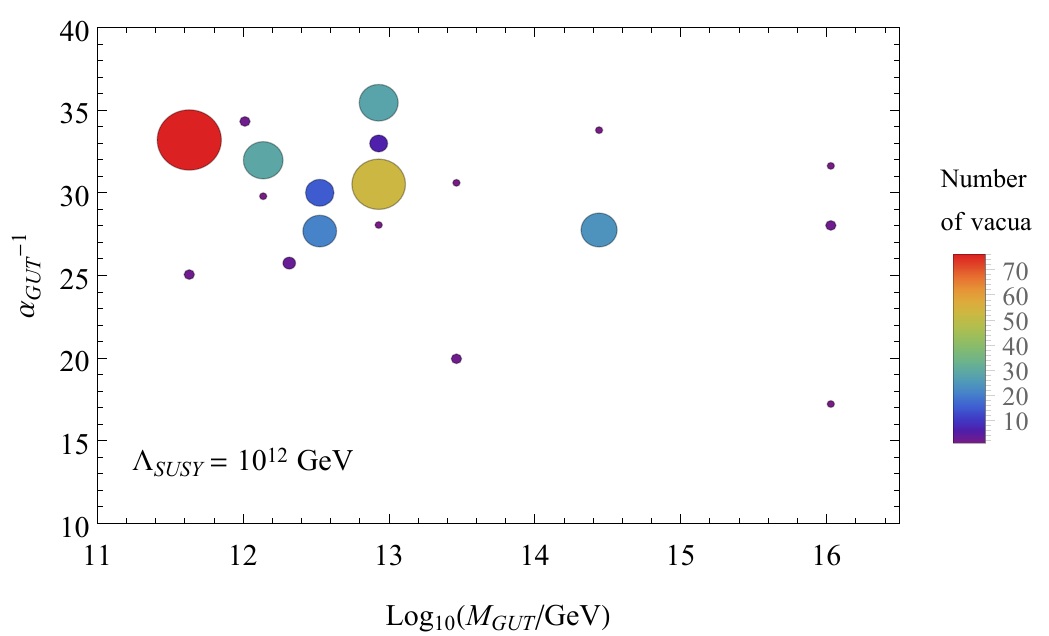}}

\subfigure{\includegraphics[width=0.49\textwidth]
{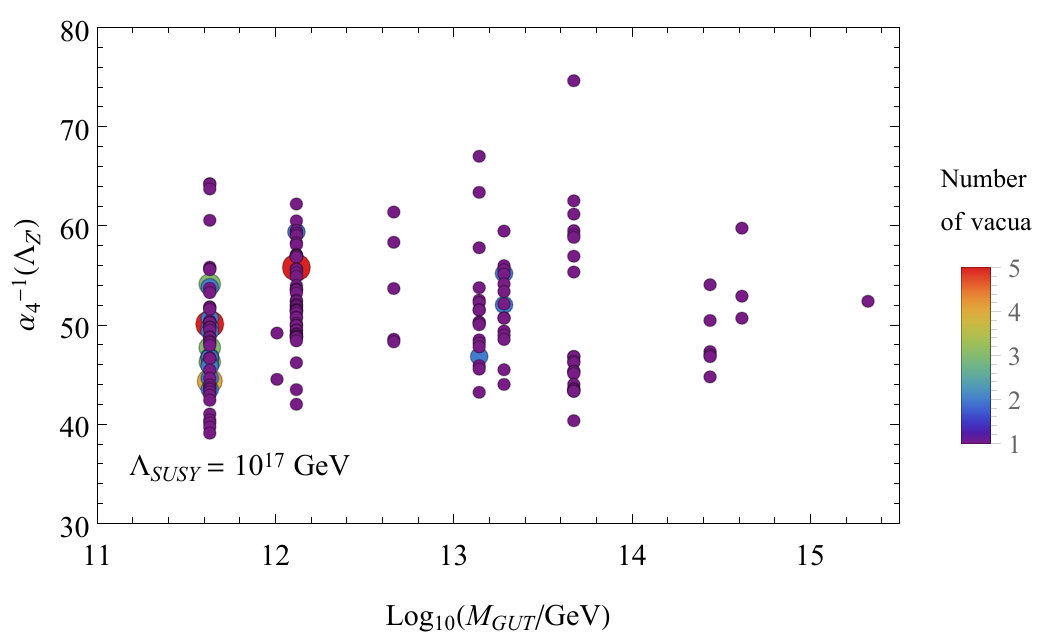}}
\subfigure{\includegraphics[width=0.49\textwidth]
{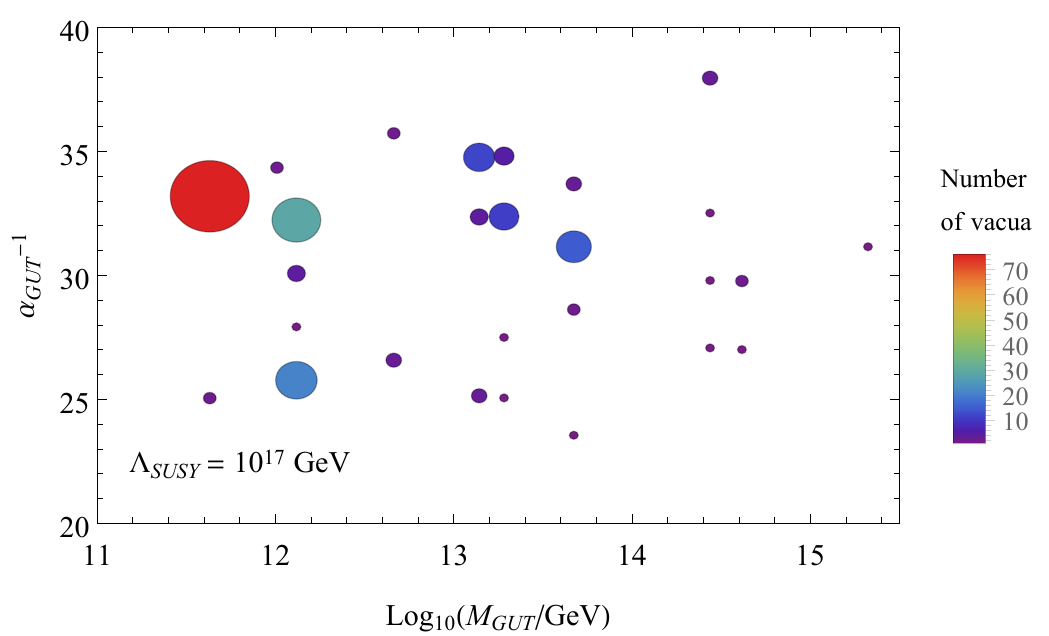}}

\caption{Vacua with different values of $M_{GUT}$, $\alpha_4^{-1}(\Lambda_{Z'})$ and $\alpha_{GUT}^{-1}$
         for three choices of $\Lambda_{SUSY}$ and partial unification, $\alpha_{GUT}\equiv\alpha_1=\alpha_2\approx\alpha_3$. 
         In the left panels, the bubbles in different colors and sizes indicate the number of vacua
         with the given values of $\alpha_4(\Lambda_{Z'})$ and $M_{GUT}$ at their center. Analogously,
         the right panels count vacua with different values of $\alpha_{GUT}$ and $M_{GUT}$.} 
\label{fig:Z8allgeom-unif}
\end{figure}

%%%%%%%%%%%%%%%% MAIN RESULTS
In figure~\ref{fig:Z8allgeom-unif} we present our main results: the values of $\alpha_4(\Lambda_{Z'})$, $\alpha_{GUT}$
and $M_{GUT}$ in the \Z8 orbifold vacua with gauge coupling unification. As before, we present how often 
we find in our effective vacua the few admissible values of the $\U1'$ coupling strength at reachable energies, $\alpha_4(\Lambda_{Z'})$,
the scale at which all coupling strengths meet, $M_{GUT}$, and the value of the coupling
strengths when they meet, $\alpha_{GUT}=\alpha_i(M_{GUT}),i=1,2,3$. 
From the top to the bottom plots, we display these results for the SUSY scales, $\Lambda_{SUSY}=2$ TeV,
$10^{12}$ GeV and $10^{17}$ GeV. 

For low-SUSY scale, we observe that the $\U1'$ coupling strength is quite restricted by
$25 \leq \alpha_4^{-1}(2\text{ TeV}) \leq 60$, or equivalently $0.46 \leq g_4(2\text{ TeV}) \leq 0.7$;
the only allowed unification scales are $M_{GUT}\in\{10^{12}\text{ GeV},6.6\x10^{13}\text{ GeV},4.1\x10^{16}\text{ GeV}\}$;
and the coupling at unification takes only a few values, restricted by $5.6 \leq \alpha_{GUT}^{-1} \leq 21.4$.
We note that $g_4(2\text{ TeV})\approx0.6$ is the most commonly present value, just below the observed
value of the $\SU2_L$ coupling.
Additionally, most of the vacua (62 out of 77) find unification at the largest $M_{GUT}$. At that scale,
the preferred value of the GUT coupling corresponds to $\alpha_{GUT}\approx1/21$, very close to the value taken
traditionally in GUTs, $\alpha_{GUT}\approx1/25$.
As a side remark, although we have considered $\alpha_i=1$ as our perturbativity limit, a stricter bound is achieved if one demands
$g_i<1$, which would imply that the values $\alpha_{GUT}^{-1}\lesssim 11$ should be disregarded and, in turn, so should the 
vacua with the lowest GUT scale $M_{GUT}=10^{12}$~GeV in this case.

For intermediate scale SUSY breaking, the variety of \Z8 orbifold vacua with unification
is richer. However, once more, there are strong restrictions on the 
possibly observable values of the $\U1'$ couplings, set by 
$38 < \alpha_4^{-1}(2\text{ TeV})\leq 64$, or equivalently $0.44 \leq g_4(2\text{ TeV}) < 0.6$.
Since the distribution of $\U1'$ coupling values at low energies is quite uniform, its 
average value can also be of some interest: $\overline{g_4(2\text{ TeV})}\approx0.5$.
Concerning the unification scale and the coupling at those energies, we find a very compact
distribution of values with $4.3\x10^{11}\text{ GeV} \leq M_{GUT} \leq 10^{16}$ GeV and 
$17 < \alpha_{GUT}^{-1} < 36$.
Most of the vacua yield $M_{GUT}\approx 4.3\x10^{11}$ GeV and $\alpha_{GUT}\approx 1/33$.
It is interesting that the higher the SUSY breaking scale, the lower the unification scale.

An intriguing observation is that models without SUSY below $M_{str}$ emerging from heterotic
orbifold compactifications produce very similar results (roughly identical in our approximations)
to those of intermediate SUSY scale. In particular, inspecting the bottom-left panel,
we see that the range of values for $\alpha_4^{-1}(2\text{ TeV})$ coincides with the previous case,
except for an isolated vacuum, which we might ignore. As a consequence, again, $\overline{g_4}\approx0.5$
at low energies. In fact, most of the vacua of this type render {\it exactly} the same $M_{GUT}$
and $\alpha_{GUT}$ as in the previous case.

%%%%%%%%%%%%%%%%%%%%%%%%%%%%%%%%%%%%%%%%%%%%%%%%%%%%%%%%%%%%%%%%%%%%%%%%%%%%%%%%%%%%%%%%%%%%%%%%%%%%%%%%%
\section{Sample model with potentially stable Higgs vacuum}
\label{sec:sample}
%%% Model 169, vacuum 5
% Log_10(M_GUT)= 12.1387807675
% alpha_GUT^{-1}= 31.95820
% alpha_4^{-1}(Lambda_Z')= 53.57959
% alpha_3^{-1}(M_GUT)= 32.19221
% b'_i=(23/3, -1/3, -19/3, 715/108)
% b_i=(59/5, 5, -2, 731/72)
%|Delta alpha|= 0.000227455

To illustrate the features of our promising vacua with $\U1'$, let us examine in one \Z8 orbifold 
sample vacuum the potential of a $\U1'$ as a tool to solve the metastability problem of the Higgs potential. 
According to ref.~\cite{DiChiara:2014wha}, SM fermions and some extra singlets with $\U1'$ charges, subject to a series
of constraints, can ameliorate the RGE running of the relevant couplings, yielding a positive Higgs self-coupling 
at all scales.

%%%%%%%%%%%%%%%%%%%%%%%%%%%%%%%%%%%%%%%%%%%%%%%%%%%%%%%%%%%%%%%%%%%%%%%%%%%%%%%%%%%%%%%%%%%%%%%%%%%%%%%%%%%%%%%%%
The shift vector $V$ and WL $W_a$ (satisfying the constraints in table~\ref{tab:Z8WLConstraints}) 
that define the gauge embedding of a particular \Z8--II (2,1) orbifold are
\begin{eqnarray}
V   &=&\tfrac{1}{4}(-7/2, 0, 0, 0, 1/2, 1/2, 5/2, 3) (-4, -1, 0, 0, 0, 1/2, 1/2, 3), \\
W_1 &=&\tfrac{1}{4}( 1, -7, -7, -5, 2, 2, 1, -3)(-3, 3, -6, -4, 1, -3, 3, 5),\quad W_6=0\,. \nonumber
\end{eqnarray} 
The resulting gauge group reads $\maG_{4D}=\maG_{SM}\x[\U1']^6\x\SU2^6$, where one $\U1'$ is (pseudo-)anomalous.
We choose a vacuum of SM singlet VEVs, such that $\maG_{4D}\to\maG_\text{eff}$ (see eq.~\eqref{eq:Geff})
spontaneously and the (correctly normalized) hypercharge and $\U1'$ generators are given by
\begin{eqnarray}
\label{eq:U1generators}
t_1 &=& \tfrac{1}{4}(1, 5/3, 5/3, -5/3, 1, 1, 1, 1)(0, 0, 0, 0, 0, 0, 0, 0), \\  %this generator had a sign wrong!
t_4 &=& \tfrac{1}{12\sqrt{2}}(-3, 0, 0, 0, 1, 1, 1, -2)(0, 0, 0, 0, 0, 8, 8, 0)\,.\nonumber
\end{eqnarray}
%%%%%%%%%%%%%%%%%%%%%%%%%%%%%%%%%%%%%%%%%%%%%%%%%%%%%%%%%%%%%%%%%%%%%%%%%%%%%%%%%%%%%%%%%%%%%%%%%%%%%%%%%%%%%%%%%%%%%%
The spectrum of the chosen vacuum, after the decoupling of vectorlike exotics w.r.t. $\maG_\text{eff}$, is displayed in 
table~\ref{spectrum}, considering scales below $\Lambda_{SUSY}=10^{12}$ GeV.
It contains the SM particles, an extra Higgs boson, few fermionic exotics and some SM singlets, which are mostly fermions. 
We choose as scalars two (instead of one) SM singlets, $s_1$ and $s_2$, to trigger the spontaneous breakdown of $\U1'$
and facilitate the decoupling of SM exotics.

%%%%%%%%%%%%%%%%%%%%%%%%%%%%%%%%%%%%%%%%%%%%%%%%%%%%%%%%%%%%%%%%%%%%%%%%%%%%
\begin{table}[!t!]
	\begin{center}
		{\footnotesize
			\begin{tabular}{|rlc|c|rlc|c|rlc|}
				\cline{1-3}\cline{5-7} \cline{9-11}
				\#  & Fermionic irrep   & Label &&  \#  & Fermionic irrep & Label &&  \#  & Scalar irrep & Label \\
				\cline{1-3}\cline{5-7}\cline{9-11}
				
				2 & $(\bs1, \bs2)_{(-1/2,\,-7/12\sqrt{2})}\phantom{A^{A^A}}$ & $\ell_{1,2}$ 
				&& 1 & $(\bsb3, \bs1)_{(1/3,\,3/4\sqrt{2})}$                 & $\bar{x}_i$ 
				&&1 & $(\bs1, \bs2)_{(1/2,\,-1/3\sqrt{2})}$                  & $H_u$ \\

				1 & $(\bs1, \bs2)_{(-1/2,\,1/3\sqrt{2})}$                    & $\ell_3$ 
				&& 1 &  $(\bs3, \bs1)_{(-1/3,\,1/12\sqrt{2})}$               & $x_i$ 
				&&1 & $(\bs1, \bs2)_{(-1/2,\,1/12\sqrt{2})}$                 & $H_d$  \\
				
				%\cline{5-7} \cline{9-11}
				2 & $(\bs1, \bs1)_{(1,\,-1/6\sqrt{2})}$                      & $\bar{e}_{1,2}$
				&& 8 &  $(\bs1, \bs2)_{(0,\,1/6\sqrt{2})}$                   & $\eta_i$
				&& 1 & $(\bs1, \bs1)_{(0,\,-1/4\sqrt{2})}$                   & $s_1$ \\
				%\cline{5-7}
				1 & $(\bs1, \bs1)_{(1,\,1/12\sqrt{2})}$                      & $\bar{e}_3$ 
				&& 8 & $(\bs1, \bs1)_{(1/2,\,1/6\sqrt{2})}$                  & $\zeta_i$ 
				&& 1 & $(\bs1, \bs1)_{(0,\,1/3\sqrt{2})}$                    & $s_2$ \\
				
				2 & $(\bs3, \bs2)_{(1/6,\,-1/6\sqrt{2})}$                    & $q_{1,2}$  
				&& 8 & $(\bs1, \bs1)_{(-1/2,\,7/12\sqrt{2})}$                & $\bar\zeta_i$  
				&&  &    &   \\
				
				1 & $(\bs3, \bs2)_{(1/6,\,1/4\sqrt{2})}$                     & $q_3$ 
				&& 8 & $(\bs1, \bs1)_{(-1/2,\,-1/12\sqrt{2})}$               & $\bar\kappa_i$  
				&&  &  &  \\
				
				2 & $(\bsb3, \bs1)_{(-2/3,\,-1/6\sqrt{2})}$                  & $\bar{u}_{1,2}$
				&& 8 & $(\bs1, \bs1)_{(1/2,\,-1/6\sqrt{2})}$                 & $\kappa_i$    
				&& & & \\
				\cline{5-7}
				1 & $(\bsb3, \bs1)_{(-2/3,\,1/12\sqrt{2})}$                  & $\bar{u}_3$  
				&& 11 & $(\bs1, \bs1)_{(0,\,1/3\sqrt{2})}$                   & $N^a_i$ 
				&&  &  &  \\
				
				2 & $(\bsb3,\bs1)_{(1/3,\,-7/12\sqrt{2})}$                   & $\bar{d}_{1,2}$ 
				&& 10 & $(\bs1, \bs1)_{(0,\,-2/3\sqrt{2})}$                  & $N^b_i$    
				&& & & \\
				
				1 & $(\bsb3, \bs1)_{(1/3,\,3/4\sqrt{2})}$                    & $\bar{d}_3$
				&& 8 & $(\bs1, \bs1)_{(0,\,-1/12\sqrt{2})}$                  & $N^c_i$
				&& & & \\
				
				&   &  
				&& 6 & $(\bs1, \bs1)_{(0,\,-5/12\sqrt{2})}$                  & $N^d_i$   
				&& & & \\
				
				&    &  
				&& 4 & $(\bs1, \bs1)_{(0,\,7/12\sqrt{2})}$                  & $N^e_i$
				&& & & \\
				
				&   &  
				&& 2 & $(\bs1, \bs1)_{(0,\,-1/4\sqrt{2})}$                  & $N^f_i$
				&& & &\\
				
				\cline{1-3}  
				\cline{5-7}
				\cline{9-11}
			\end{tabular}
			\caption{Massless spectrum for a vacuum with gauge group $\maG_\text{eff}$.
				Representations  with respect to $\SU3_c \x \SU2_L$ are given  in  bold face,
				the hypercharge and the $\U1'$ charge are indicated as subscript. The frame on the 
				left corresponds to the SM fermions, the middle frame to fermionic exotics, 
				and the right frame shows scalars including the Higgs fields.}
			\label{spectrum}
		}
	\end{center}
\end{table}
%%%%%%%%%%%%%%%%%%%%%%%%%%%%%%%%%%%%%%%%%%%%%%%%%%%%%%%%%%%%%%%%%%%%%%%%%%%%

\begin{figure}[t!]
	\includegraphics[width=0.65\textwidth]{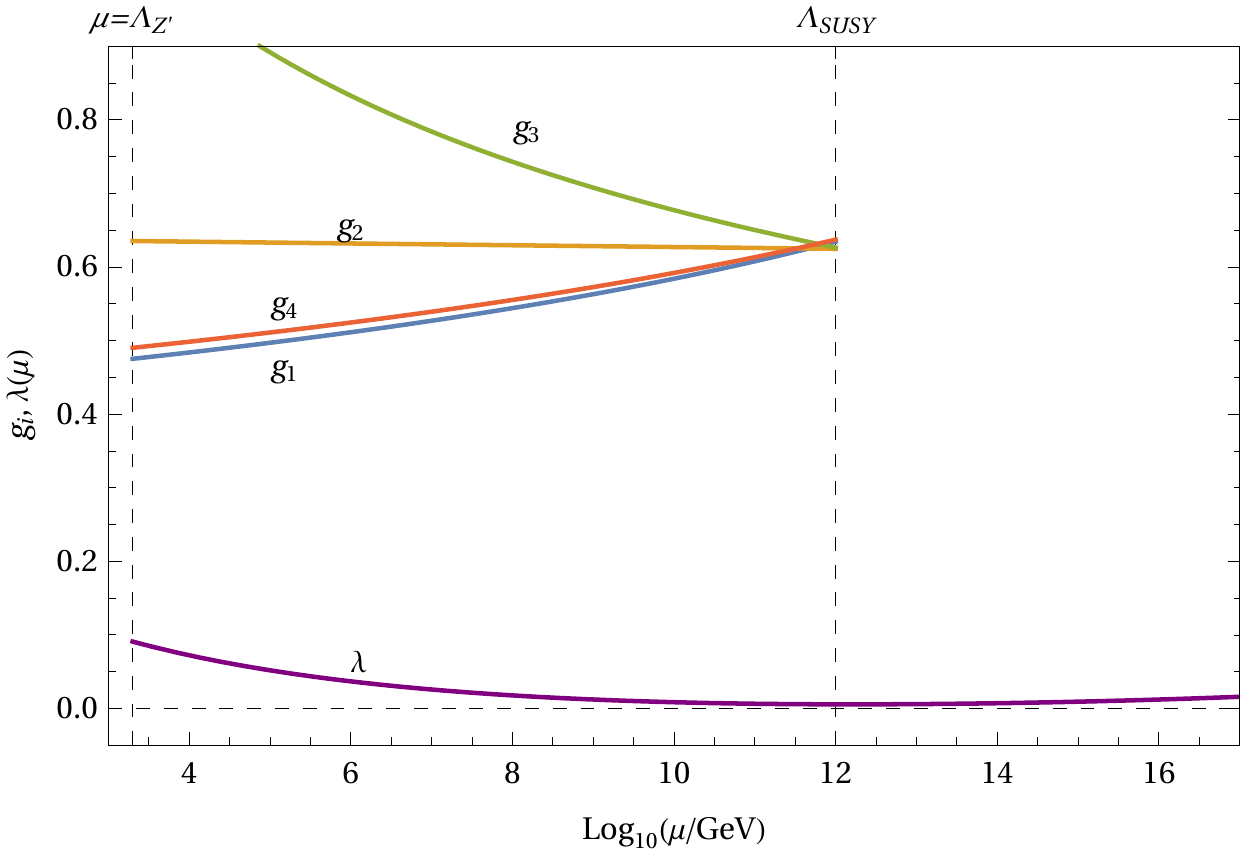}
	\centering
	\caption{RGE running of the quartic Higgs coupling and the gauge couplings of $\U1_Y,\SU2_L,\SU3_C$ and $\U1'$ in a \Z8--II (2,1) sample vacuum.
		The gauge couplings meet at $M_{GUT}\approx10^{12}$ GeV with a value $\alpha_{GUT}\approx1/32$. 
		We see that at $\Lambda_{Z'}=2$ TeV, the $\U1'$ coupling has the value $g_4\approx0.49$. From this plot, 
		we observe that the quartic Higgs coupling remains perturbative and positive, yielding a plausible solution 
		of the vacuum meta-stability problem of the SM.
		}
	\label{fig:stabhiggs}
\end{figure}

We compute the RGE running of the couplings in this model by using SARAH~\cite{Staub:2013tta}.
First, by applying our approach, we find $g_4(2\text{ TeV})\approx0.49$ and a unification scale 
$M_{GUT}\sim\Lambda_{SUSY}=10^{12}$ GeV. Supposing that the scalar fields $s_1$ and $s_2$ develop
VEVs, we note that the fermionic exotics exhibit couplings that allow them to be decoupled below $\Lambda_{Z'}=2$ TeV
while the $\U1'$ is spontaneously broken, so that we can consider only the SM spectrum below $\Lambda_{Z'}$.
Taking $g_1=0.3587$, $g_2=0.6482$, $g_3=1.1645$, the top Yukawa $Y^u_{33}=0.9356$ and the quartic Higgs 
self-coupling $\lambda=0.127$ at the top-mass scale $m_t=173.1$ GeV (see e.g.~\cite{Khan:2014kba}),
we let the SM couplings evolve below $\Lambda_{Z'}$. For $\Lambda_{Z'}<\mu<\Lambda_{SUSY}$, we include all exotics
of table~\ref{spectrum} and further suppose that $H_u$ dominates the quartic Higgs 
self-coupling in order to carefully study the evolution of that coupling. Our findings are shown in
fig.~\ref{fig:stabhiggs}, where we have extended our description of $\lambda$ above $\Lambda_{SUSY}$
to make sure that perturbativity is not lost. In order to test the strength of our study,
we have also allowed for non-trivial values of other quartic couplings (those of $H_d$, $s_1$ and $s_2$)
and found that our result is not altered as long as those couplings are taken
close to the value of $\lambda$ at $m_t$. Thus, it is possible to state that, although our model
differs from those of ref.~\cite{DiChiara:2014wha}, our charges and $\U1'$ coupling constant combine 
together to yield a stable Higgs vacuum, as in their cases.

This model admits further interesting phenomenology. Let us roughly explore 
here some aspects concerning the fermion masses in this model. 
Based on the compactification scheme, the dominant contributions to the mass terms in the effective Lagrangian are 
given by
\begin{align*}
\begin{split}
{\small
\mathcal{L} \supset -Y^{u}_{33}\bar{u}_3H_u^{\dagger}q_3-Y^{u}_{11,22}\bar{u}_{1,2}H_u^{\dagger}q_{1,2}s_2^2
- Y^{d}_{33}\bar{d}_3H_d^{\dagger}q_3 s^7_1s_2^2
- Y^{d}_{11,22}\bar{d}_{1,2}H_d^{\dagger}q_{1,2} s_2^2
}
\\
{\small
- Y^{\ell}_{33}\bar{e}_3H_d^{\dagger}\ell_{3}s^2_1
- Y^{\ell}_{11,22}\bar{e}_{1,2}H_d^{\dagger}\ell_{1,2}s_2^2
- Y^{\nu}_{ii}N_i^bH_u^{\dagger}\ell_{3} s_2^2
- k_{ij} N_i^aN_j^cs_1+h.c., 
}
\end{split}
\end{align*}
where we suppose that the singlet VEVs can be chosen allowing some tuning. For example, we find, at this level, 
that the top quark has the largest mass as the corresponding Yukawa coupling appears unsupressed, allowed by
all symmetries of the string construction. Other Yukawas are suppressed by the singlet VEVs. For example, 
if $\U1'$ is spontaneously broken such that
$\langle s_2\rangle ^2 \sim \mathcal{O}(10^{-5})$, 
$\langle s_1\rangle \sim \mathcal{O}(10)$, and
$\langle H_d \rangle \sim \mathcal{O}(10^{-4})\langle H_u \rangle$, one arrives at the correct relations
$m_t/m_u\approx10^{5}$, $m_{t}/m_{b}\approx 10^{2}$ and $m_{t}/m_{\tau}\approx 10^{2}$, where all 
coefficients $Y^{u,d,\ell}$ are (unsupressed) of order unity because untwisted fields
appear in each coupling. Additionally, we observe that neutrino masses of the right order are generated 
through a type-I see-saw mechanism, where $N_i$ are heavy right-handed neutrinos. 
Further, while $\U1'$ breaks down, the exotics of the middle frame of table~\ref{spectrum} also 
develop masses of order $\langle s_{1,2}\rangle$ and could also be detected as a signal of
this kind of models.
On the less bright side, in our model the electron and down quark are very light, $m_t/m_{e}=m_{t}/m_{d}\approx 10^{9}$,
the chosen VEVs require large fine-tuning because the effective theory is defined 
at $\Lambda_{SUSY}$, and there is a residual flavor symmetry between
the first and second generation.  We expect that a more careful analysis of  
additional details of the model, such as the SUSY and flavor breakdown, shall 
provide solutions to these issues, but this analysis is beyond the scope of this letter.

%%%%%%%%%%%%%%%%%%%%%%%%%%%%%%%%%%%%%%%%%%%%%%%%%%%%%%%%%%%%%%%%%%%%%%%%%%%%%%%%%%%%%%%%%%%%%%%%%%%%%%%%%
\section{Final remarks and outlook}
\label{sec:conclusions}

By means of the 1-loop RGE,
we have systematically studied the TeV-scale value of the $\U1'$ coupling constant
in vacua arising from \Z8 heterotic orbifold compactifications whose matter content
exhibits the MSSM spectrum plus vectorlike exotics at the string scale. We have 
restricted ourselves to vacua with only one non-anomalous $\U1'$ gauge symmetry, and whose SM gauge couplings
have the observed values and unify at a model-dependent GUT scale, below
the string scale. Only between 0.5\% and 1.5\% of all possible vacua satisfy these conditions.

Supposing that the $\U1'$ breakdown scale is of order of few TeV, reachable at colliders,
we find that for TeV SUSY the $\U1'$ coupling constant is restricted
in our constructions to lie in the small range $0.46<g_4<0.7$. This range is 
further reduced to $0.44<g_4<0.6$ if one allows SUSY to be broken
at a scale larger than $10^{12}$ GeV. Models with such couplings
exhibit exotic fermions, in addition to a multi-TeV $Z'$, that may be
detected soon.

We have found that also the unification scale is restricted in 
\Z8 orbifold vacua to be roughly either $10^{14}$ GeV or $10^{16}$ GeV
for low-scale SUSY, or preferably about $10^{12}$ GeV for intermediate
SUSY breaking scale or higher.

We have also studied the properties of a sample model, finding that,
if intermediate scale SUSY is realized, there are \Z8 orbifold vacua 
that may be furnished with the ingredients to stabilize the Higgs vacuum.
The details of such vacua and mechanism are left for future work.

In our scheme, the dynamics of the spontaneous breaking of $\U1'$ requires
large fine-tuning to establish the hierarchies $\Lambda_{Z'}\ll \Lambda_{SUSY}\ll M_{str}$.
In a model-dependent basis, it could however be possible that the potential of
SM singlets and gaugino condensates conspire to yield such hierarchies. One may also
wonder whether this simplifies in non-supersymmetric heterotic orbifolds.
Another issue is the details of the RGE at the SUSY breaking scale, including
the decoupling of superpartners, which may require a treatment such as in~\cite{Beneke:2008wj}.
These important questions shall be the goal of future projects.

%%%%%%%%%%%%%%%%%%%%%%%%%%%%%%%%%%%%%%%%%%%%%%%%%%%%%%%%%%%%%%%%%%%%%%%%%%
\paragraph{Acknowledgments.}
It is a pleasure to thank Jens Erler for very useful discussions and motivation to pursue this work.
SR-S would like to thank Rafael Alapisco-Ar\'ambula, who participated in an early stage of this project.
SR-S is grateful to the Bethe Center for Theoretical Physics and the Mainz Institute for Theoretical
Physics for the hospitality.
This work was partly supported by DGAPA-PAPIIT grant IN100217 and CONACyT grants F-252167
and 278017.

%%%%%%%%%%%%%%%%%%%%%%%%%%%%%%%%%%%%%%%%%%%%%%%%%%%%%%%%%%%%%%%%%%%%%%%%%%
%  Bibliography
%%%%%%%%%%%%%%%%%%%%%%%%%%%%%%%%%%%%%%%%%%%%%%%%%%%%%%%%%%%%%%%%%%%%%%%%%%
%\bibliography{Orbifold}
%\bibliographystyle{NewArXiv}

\providecommand{\bysame}{\leavevmode\hbox to3em{\hrulefill}\thinspace}
\frenchspacing
\newcommand{\origttfamily}{}
\let\origttfamily=\ttfamily
\renewcommand{\ttfamily}{\origttfamily \hyphenchar\font=`\-}

\end{document}